\documentclass[10pt, conference, compsocconf]{IEEEtran}
\usepackage{cite}
\ifCLASSINFOpdf
  \usepackage[pdftex]{graphicx}
\else
  \usepackage[dvips]{graphicx}
\fi
\usepackage[cmex10]{amsmath}
\usepackage{array}
\usepackage{theorem}
\usepackage[caption=false,font=footnotesize]{subfig}
\usepackage{fixltx2e}
\usepackage{url}
\usepackage{acronym}
\usepackage{comment} 

{\theorembodyfont{\rmfamily}
   }
{\theorembodyfont{\rmfamily}
   }
{\theorembodyfont{\rmfamily}
   }
{\theorembodyfont{\rmfamily}
   }

\acrodef{LDPC}{low-density parity-check}
\acrodef{MDPC}{moderate-density parity-check}
\acrodef{QC}{quasi-cyclic}
\acrodef{QC-LDPC}{quasi-cyclic low-density parity-check}
\acrodef{QC-MDPC}{quasi-cyclic moderate-density parity-check}
\acrodef{RSA}{Rivest, Shamir, Adleman}
\acrodef{BF}{bit flipping}
\acrodef{SPA}{sum product algorithm}
\acrodef{RDF}{random difference families}
\acrodef{ISDA}{information set decoding attacks}
\acrodef{DCA}{dual code attacks}
\acrodef{WF}{work factor}
\acrodef{BER}{bit error rate}
\acrodef{CER}{codeword error rate}
\acrodef{BSC}{binary symmetric channel}

\begin{document}

\title{Improving the efficiency of the LDPC code-based \\ McEliece cryptosystem through irregular codes}

\author{\IEEEauthorblockN{Marco Baldi, Marco Bianchi, Nicola Maturo, Franco Chiaraluce,\\}
\IEEEauthorblockA{DII, Universit\`a Politecnica delle Marche,\\
Ancona, Italy\\
Email: \{m.baldi, m.bianchi, n.maturo, f.chiaraluce\}@univpm.it}}

\maketitle

\begin{abstract}
We consider the framework of the McEliece cryptosystem based on \ac{LDPC} codes, which is a promising
post-quantum alternative to classical public key cryptosystems.
The use of LDPC codes in this context allows to achieve good security levels with very compact keys, which
is an important advantage over the classical McEliece cryptosystem based on Goppa codes.
However, only regular \ac{LDPC} codes have been considered up to now, while some further improvement can be
achieved by using irregular \ac{LDPC} codes, which are known to achieve better error correction performance
than regular \ac{LDPC} codes.
This is shown in this paper, for the first time at our knowledge.
The possible use of irregular transformation matrices is also investigated, which further increases the efficiency
of the system, especially in regard to the public key size.
\end{abstract}


\section{Introduction}
\label{sec:Intro}
\let\thefootnote\relax\footnotetext{This work was supported in part by the MIUR project ``ESCAPADE''
(Grant RBFR105NLC) under the ``FIRB -– Futuro in Ricerca 2010'' funding program.}

A renewed interest is being devoted to code-based cryptosystems, since they are recognized to be
able to resist attacks based on quantum computers, which will seriously endanger widespread
solutions, like the \ac{RSA} system, based on the integer factorization problem.
The best known code-based public key cryptosystem is the McEliece cryptosystem \cite{McEliece1978},
which relies on the problem of decoding a random linear block code with no visible structure.
This system, in its original formulation, has never encountered any polynomial time attack, and
is able to guarantee very fast encryption and decryption procedures.
Its major drawbacks are the encryption rate, which is smaller than $1$, and, most of all,
the large size of its public keys.

In fact, the original solution adopts Goppa codes, which are able to ensure very high security levels,
but require the public matrices to be unstructured. Hence, their storage needs a great amount of memory.
The most recent proposals concerning Goppa codes provide updated choices of the system parameters 
to reduce the public key size and increase the security level \cite{Bernstein2008}.
Despite this, the public key size is still very large: the parameters proposed in \cite{Bernstein2008}
to achieve $128$-bit security yield a public key size of $1,537,536$ bits.

Replacing Goppa codes with structured codes allows achieving considerable reductions in the key size,
though security issues must be taken into account.
Recently, several proposals based on \ac{QC-LDPC} codes have appeared \cite{Baldi2008, Baldi2012, Baldi2013, Misoczki2012, Biasi2012},
showing that these codes are actually a promising alternative to traditional Goppa codes.
\ac{LDPC} codes are capacity achieving codes \cite{Richardson2001} defined through sparse parity-check matrices.
They are employed in several frameworks \cite{Paolini2009, Baldi2012b, Baldi2012c}, and also used in some security-related contexts \cite{Baldi2012a}.
Their use in the McEliece cryptosystem has been studied since several years \cite{Monico2000, Baldi2007ISIT, Baldi2007ICC},
and some refinements have been progressively introduced to eventually achieve a secure instance of the system.
The use of structured \ac{LDPC} codes, like \ac{QC-LDPC} codes, allows to considerably reduce the key size,
though renouncing the sparse character of the public matrices.

Up to now only regular \ac{QC-LDPC} codes have been considered for the use in this context, while it is well
known that irregular \ac{LDPC} codes can achieve better error correcting performance than regular codes \cite{Luby2001a}.
In this paper, we show that such feature allows to further reduce the public key size.
Another contribution to the same goal comes from the adoption of irregular transformation matrices.

The paper is organized as follows:
in Section \ref{sec:System}, we recall the \ac{QC-LDPC}-code based McEliece cryptosystem;
in Section \ref{sec:IrrCodes}, we define irregular \ac{QC-LDPC} codes to be used in this context, and assess their performance through simulations;
in Section \ref{sec:Threshold}, we provide a theoretical tool for estimating the error correction capability of irregular codes;
in Section \ref{sec:SecLevel}, we assess the security level of the system;
in Section \ref{sec:Examples}, we show the advantage achieved by irregular codes in terms of key size through some examples;
finally, Section \ref{sec:Conclusion} concludes the paper.

\section{\ac{QC-LDPC} code-based McEliece cryptosystem}
\label{sec:System}

The McEliece cryptosystem based on \ac{QC-LDPC} codes uses codes with length $n = n_0 \cdot p$, dimension 
$k = k_0 \cdot p$ and redundancy $r = p$, where $n_0$ is a small integer (\textit{e.g.}, $n_0 = 2,3,4$), $k_0 = n_0-1$, and $p$
is a large integer (on the order of some thousands or more).
It follows that the code rate is $\frac{n_0-1}{n_0}$, which coincides with the encryption rate.
Differently from other solutions, like \ac{RSA}, the McEliece cryptosystem has encryption rate $< 1$,
which yields some overhead in the ciphertext, due to the code redundancy.
While the original McEliece cryptosystem used codes with rate about $1/2$, its most recent variants
are focused on code rates on the order of $0.7$.
Concerning the \ac{QC-LDPC} code-based variant, we consider the choice $n_0=4$, that is, an
encryption rate equal to $3/4$, which is in line with the most recent proposals.

In the \ac{QC-LDPC} code-based McEliece cryptosystem, the main component of the private key 
is a \ac{QC-LDPC} matrix having the following form \cite{Baldi2011a, Baldi2012}:
\begin{equation}
\mathbf{H} = \left[ \mathbf{H}_{0} | \mathbf{H}_{1} | \ldots |\mathbf{H}_{n_0-1} \right],
\label{eq:HCircRow}
\end{equation}
where each $\mathbf{H}_{i}$ is a circulant matrix with size $p \times p$.

In all previous proposals, the matrix $\mathbf{H}$ as in \eqref{eq:HCircRow} was regular,
that is, with constant column weight $d_v$ and constant row weight $d_c = n_0 d_v$.
In this work, we analyze a more general form of $\mathbf{H}$ by considering non-constant
column weights.
However, differently from completely general irregular \ac{LDPC} codes, we must preserve 
the \ac{QC} nature of the codes, since it provides important advantages in terms of the public 
key size.

Hence, we consider a private \ac{QC-LDPC} matrix which still has the form \eqref{eq:HCircRow},
but formed by $n_0$ circulant blocks with different column weights: $\left\{d_v^{(0)}, d_v^{(1)}, d_v^{(2)}, \ldots, d_v^{(n_0-1)} \right\}$.
Now, $d_v$ has the meaning of average column weight, i.e., $d_v = \sum_{i=0}^{n_0-1} d_v^{(i)} / n_0$, and
the row weight is still constant and equal to $d_c = n_0 d_v$.

Other two matrices are needed to form the private key: a $k \times k$ non singular random scrambling matrix $\mathbf{S}$ 
and an $n \times n$ non singular sparse transformation matrix $\mathbf{Q}$.
In the previous versions of the \ac{QC-LDPC} code-based McEliece cryptosystem \cite{Baldi2008, Baldi2012}, also $\mathbf{Q}$ 
was a regular matrix, with fixed row and column weight $m$.
We generalize it by defining $\mathbf{Q}$ as a sparse irregular matrix with average row and column
weight $m$. As we will see in the following, $\mathbf{Q}$ affects the weight of the error vectors.
In order to maintain its effect uniform, independently of the error vector, we impose that the row and column weights
of $\mathbf{Q}$ have minimal dispersion around their mean, that is, they differ from $m$ by less than $1$.
This allows choosing rational values for $m$, which gives a further degree of freedom for improving the
system efficiency.
Moreover, in order to preserve the \ac{QC} structure for the public matrices, the matrix $\mathbf{Q}$ must be
\ac{QC} as well, that is, formed by $n_0 \times n_0$ circulant sub-matrices, each with size $p \times p$.
This choice limits the resolution on the value of $m$, which cannot vary by less than $1/n_0^2$, but it is sufficient
to ensure enough granularity in this context.
For preserving the \ac{QC} form of the public keys, also $\mathbf{S}$ must be \ac{QC}, that is, formed by $k_0 \times k_0$ 
circulant blocks with size $p \times p$.

The public key is obtained as $\mathbf{G}' = \mathbf{S}^{-1} \cdot \mathbf{G} \cdot{\mathbf{Q}^{-1}}$;
hence, its size depends on the representation of $\mathbf{G}'$.
The \ac{QC} nature of the codes and of the scrambling and transformation matrices allows to achieve a very compact
representation, since each circulant block is simply described by its first row.
In addition, using a CCA2 secure conversion of the system \cite{Bernstein2008} allows adopting public matrices
in systematic form; hence, the public key size becomes $k_0 \cdot (n_0 - k_0) \cdot p = (n_0 - 1) \cdot p$ bits.
This gives an important improvement with respect to Goppa code-based instances.

Similarly to the original McEliece cryptosystem, encryption is performed according to the following steps:
\begin{enumerate}
\item Alice gets Bob's public key $\mathbf{G}'$.
\item She divides her message into $k$-bit vectors.
\item For each $k$-bit vector $\mathbf{u}$, she generates a random intentional error vector $\mathbf{e}$ with weight $t'$.
\item She encrypts $\mathbf{u}$ into $\mathbf{x}$ as follows:
\begin{equation}
\mathbf{x} = \mathbf{u} \cdot \mathbf{G}' + \mathbf{e}.
\end{equation}
\end{enumerate}

Decryption is performed as follows:
\begin{enumerate}
\item Bob inverts the secret transformation:
\begin{equation}
\mathbf{x}' = \mathbf{x} \cdot \mathbf{Q} = \mathbf{u} \cdot \mathbf{S}^{-1} \cdot \mathbf{G} + \mathbf{e} \cdot \mathbf{Q}
\end{equation}
and obtains a codeword of the secret LDPC code affected by
the error vector $\mathbf{e} \cdot \mathbf{Q}$, with weight $\leq t=t'm$.
\item He corrects all the errors through LDPC decoding and obtains $\mathbf{u} \cdot \mathbf{S}^{-1}$.
\item He recovers $\mathbf{u}$ through multiplication by $\mathbf{S}$.
\end{enumerate}

The main difference with respect to the original McEliece cryptosystem is
in the matrix $\mathbf{Q}$, which was a permutation matrix in the original
system, while now it has average row and column weight $m > 1$.
This causes propagation of the intentional errors during decryption, and
their number is increased at most by a factor $m$.
Hence, the secret \ac{QC-LDPC} code must be able to correct up to $t = t'm$ errors, rather than $t'$.
On the other hand, this allows protecting the private key from attacks aimed at exploiting
its sparsity, as we will briefly recall in the following.

Moreover, the sparse parity-check matrix of the secret code ($\mathbf{H}$) is mapped into a new matrix for the public code:
\begin{equation}
\mathbf{H'} = \mathbf{H} \cdot \mathbf{Q}^T
\end{equation}
(superscript $T$ denotes the transpose) and, though a suitable choice of $m$, the density of $\mathbf{H'}$ can be made high enough to avoid attacks to the dual code (see Section \ref{sec:SecLevel}).

\section{Irregular \ac{QC-LDPC} codes performance}
\label{sec:IrrCodes}

It is known that irregular \ac{LDPC} codes are able to achieve better performance than 
regular ones \cite{Luby2001a}.
Looking at the code Tanner graph, an irregular code is defined through its variable and 
check nodes degree distributions.
According to the notation in \cite{Luby2001a}, an irregular Tanner graph with maximum 
variable node degree $\overline{d_{v} }$ and maximum check node degree $\overline{d_{c} }$
is described through two sequences, ($\lambda _{1} ,\ldots ,\lambda _{\overline{d_{v}}} $) and 
($\rho _{1} ,\ldots ,\rho _{\overline{d_{c}}} $), such that $\lambda _{i} $ ($\rho _{i} $) is 
the fraction of edges connected to variable (check) nodes with degree $i$.
These sequences can be used as the coefficients of two polynomials, $\lambda(x)$ and
$\rho(x)$, describing the edge degree distributions:

\begin{equation} \label{eq:EdgePersp} 
\left\{\begin{array}{c} {\lambda \left(x\right)=\sum _{i=1}^{\overline{d_{v} }}\lambda _{i} x^{i-1},} \\
{\rho \left(x\right)=\sum _{i=1}^{\overline{d_{c} }}\rho _{i} x^{i-1}  .} \end{array}\right.  
\end{equation} 

$\lambda(x)$ and $\rho(x)$ describe the code degree distributions from the edge perspective.
Alternatively, the same distributions can be described from the node perspective, through
two other polynomials, $v(x)$ and $c(x)$.
Their coefficients, noted by $v_{i}$ and $c_{i}$, are computed as the fractions of variable and check
nodes with degree $i$.
$\lambda(x)$ and $\rho(x)$ can be translated into $v(x)$ and $c(x)$ as follows \cite{Johnson2010}:
\begin{equation} \label{eq:NodePersp} 
\left\{\begin{array}{c} {v_{i} =\frac{{\lambda _{i} \mathord{\left/ {\vphantom {\lambda _{i}  i}} \right. \kern-\nulldelimiterspace} i} }{\sum _{j=1}^{\overline{d_{v} }}{\lambda _{j} \mathord{\left/ {\vphantom {\lambda _{j}  j}} \right. \kern-\nulldelimiterspace} j}  } ,} \\
{c_{i} =\frac{{\rho _{i} \mathord{\left/ {\vphantom {\rho _{i}  i}} \right. \kern-\nulldelimiterspace} i} }{\sum _{j=1}^{\overline{d_{c} }}{\rho _{j} \mathord{\left/ {\vphantom {\rho _{j}  j}} \right. \kern-\nulldelimiterspace} j}  } .} \end{array}\right.  
\end{equation}

According to our choices, all check nodes have the same degree $d_c$, whereas, because of (\ref{eq:HCircRow}) and the assumption of a (possible) different column weight for each circulant block, a fraction $1/n_0$ of the variable nodes has degree $d_v^{(i)}$, $i = 0, \ldots, n_0 - 1$. This yields the following simple forms for $v(x)$ and $c(x)$, that will be used in the following:
\begin{equation}
\left\{\begin{array}{rcl}
v(x) & = & \sum_{i=0}^{n_0-1} \frac{x^{d_v^{(i)}-1}}{n_0}, \\
c(x) & = & x^{d_c-1}.
\end{array}\right.
\end{equation}
We note that these polynomials do not correspond to optimized degree distributions, due to
the constraints imposed by the very special form \eqref{eq:HCircRow} for $\mathbf{H}$.
In addition, the minimum value of $d_v^{(i)}, i = 0, 1, \ldots, n_0-1$, is lower bounded
for security reasons. In fact, each circulant block $\mathbf{H}_i, i = 0, 1, \ldots, n_0-1$,
can be chosen in ${p \choose d_v^{(i)}}$ different ways, and we do not want this number
to decrease enough to allow an attacker to enumerate them.
However, despite being forced to obey such constraints, these irregular codes achieve significant error rate
performance improvements with respect to the regular codes considered up to now.

\begin{figure}[tb]
\begin{centering}
\includegraphics[keepaspectratio, width=80mm]{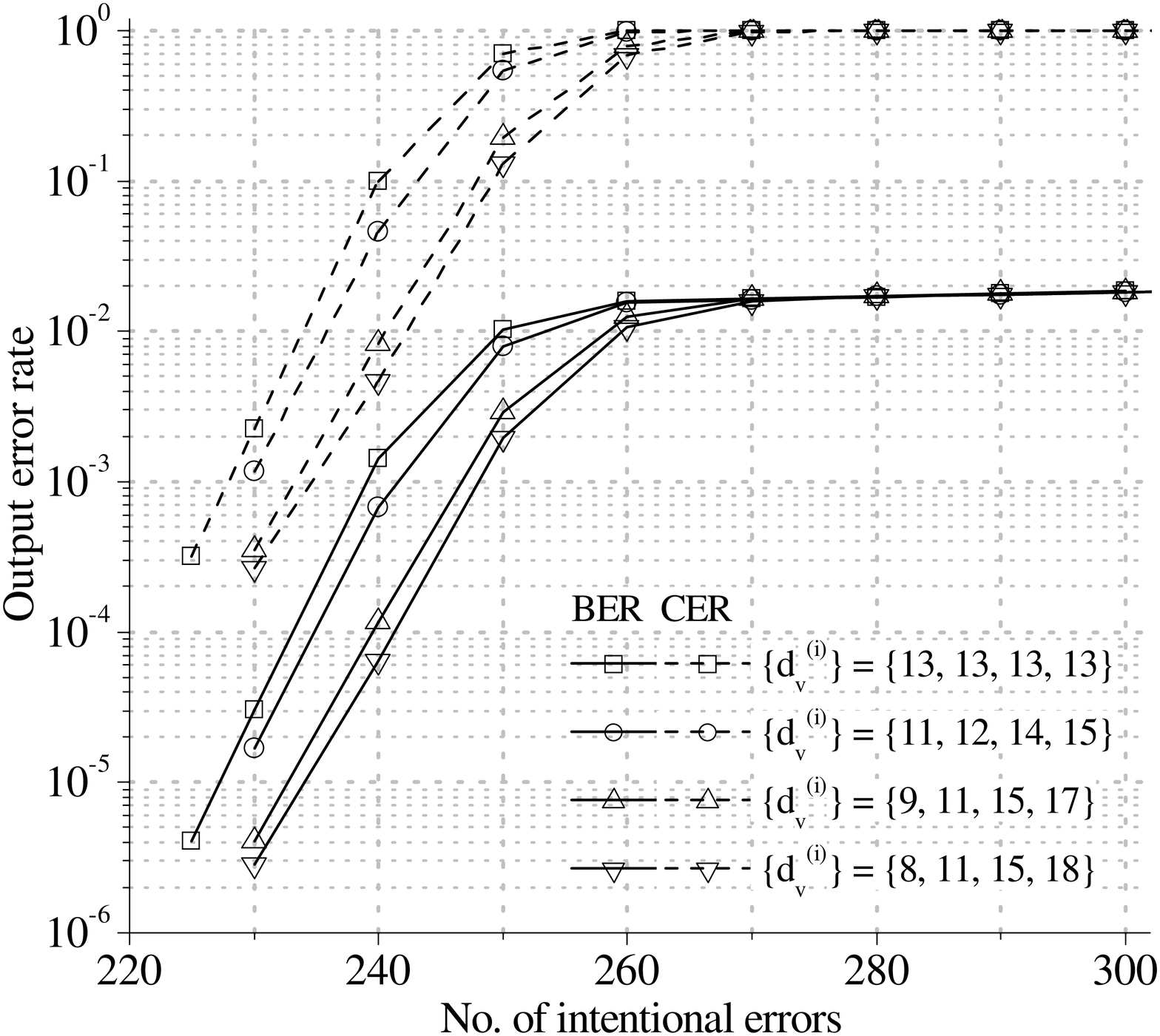}
\caption{Simulated \ac{SPA} decoding performance (\ac{BER} and \ac{CER}) for \ac{RDF}-based regular and irregular \ac{QC-LDPC} codes with $n_0=4$, $p=4096$ and $d_v=13$.}
\label{fig:Simulations}
\par\end{centering}
\end{figure}

To confirm this fact, we have focused on a fixed set of code parameters and simulated the
performance achievable by regular and irregular \ac{QC-LDPC} codes.
We have considered $n_0=4$, $p=4096$, $d_v=13$, and designed four \ac{QC-LDPC} codes
through \ac{RDF} \cite{Baldi2007ISIT}.
One of them is regular, with $d_v^{(i)} = 13$, $i = 0, \ldots, 3$, whereas the other
three are irregular, with:
\begin{itemize}
\item $d_v^{(0)} = 11, d_v^{(1)} = 12, d_v^{(2)} = 14, d_v^{(3)} = 15$,
\item $d_v^{(0)} = 9, d_v^{(1)} = 11, d_v^{(2)} = 15, d_v^{(3)} = 17$,
\item $d_v^{(0)} = 8, d_v^{(1)} = 11, d_v^{(2)} = 15, d_v^{(3)} = 18$.
\end{itemize}
Their decoding has been performed through the logarithmic version of the iterative soft-decision
\ac{SPA} \cite{Hagenauer1996}.
Performance of the \ac{SPA} is affected by quantization issues \cite{Baldi2009a},
hence we have used full precision floating point operations in our simulations.
Fig. \ref{fig:Simulations} reports the residual \ac{BER} and \ac{CER} after decoding for the
considered codes.
As expected, we observe that the irregular codes outperform the regular code
with the same parameters, and increasing the code irregularity gives a larger gain.

\section{Decoding threshold}
\label{sec:Threshold}

Despite numerical simulations provide a precise and practical assessment of the error correcting
performance of \ac{LDPC} codes, running them for each possible choice of the code parameters
and the node degree distributions is extremely time consuming.
On the other hand, no theoretical tools exist for predicting the correction capability of finite length
\ac{LDPC} codes through closed form expressions.
However, in applications which do not allow the availability of soft information from the channel, like the one here
considered, a good estimate of the correction capability can be obtained by computing the convergence
threshold of the \ac{BF} decoding algorithm.

Hence, as done in \cite{Baldi2012}, we resort to the \ac{BF} decoding threshold by extending its
computation to the case in which the codes are irregular.
The main difference with respect to the case of regular codes is in the fact that the decision
threshold $b$ is no longer unique for all variable nodes, but varies with their degree.
So, for the considered codes, up to $n_0$ different decision thresholds are used:
$b^{(i)} \leq d_{v}^{(i)}-1, i = 0, 1, \ldots, n_0-1$.
As regards other aspects, the algorithm works in the same way as for regular codes \cite{Baldi2012}.
The choice of the values $b^{(i)}$ is very important. In the original Gallager's work \cite{Gallager1963},
two algorithms were proposed: in the so-called Algorithm A, the decision thresholds are fixed to $b^{(i)} = d_{v}^{(i)}-1$,
while in the so-called Algorithm B they can vary between $\left\lceil d_{v}^{(i)}/2 \right\rceil$
and $d_{v}^{(i)}-1$ during decoding ($\left\lceil \cdot \right\rceil$ is the ceiling function). 
While Algorithm A is simpler to implement, Algorithm B is able to achieve better performance.
Both algorithms implement an iterative decision process:
i) each check node sends each neighboring variable node the binary sum of all its other neighboring
variable nodes and ii) each variable node sends each neighboring check node its initial value, flipped
or not, based on the count of unsatisfied parity-check sums coming from the other check nodes, and its
comparison with the decision threshold.
%The process of not taking into account the message coming from the current destination node is called marginalization of the messages passed through the graph.

The advantage of using \ac{BF} decoding in this context is that its decoding threshold can be estimated
through theoretical arguments very similar to those developed in \cite{Luby2001a}, that extended 
the original Gallager's probability recursion \cite{Gallager1963} to the case of irregular graphs. However, unlike \cite{Luby2001a}, where a binary symmetric channel was considered, the current scenario is equivalent to a channel able to introduce a fixed number of errors in each transmitted vector.
For the \ac{QC-LDPC} codes introduced in Section \ref{sec:System},
the probability that, in an iteration, the message originating from a
variable node is correct can be expressed as:
\begin{equation}
f^{b}\left(j,q_l\right)=\sum^{d_v^{(j)}-1}_{z=b^{(j)}}{d_v^{(j)}-1 \choose z} {\left[p^{ic}\left(q_l\right)\right]}^z{\left[p^{ii}\left(q_l\right)\right]}^{d_v^{(j)}-1-z},
\label{eq:fb}
\end{equation}
while the probability that, in an iteration, a bit that is not in error is incorrectly evaluated is:
\begin{equation}
g^{b}\left(j,q_l\right)=\sum^{d_v^{(j)}-1}_{z=b^{(j)}}{d_v^{(j)}-1 \choose z} {\left[p^{ci}\left(q_l\right)\right]}^z{\left[p^{cc}\left(q_l\right)\right]}^{d_v^{(j)}-1-z}.
\label{eq:gb}
\end{equation}
In \eqref{eq:fb} and \eqref{eq:gb}, as in \cite{Baldi2012}, we have:
\begin{equation}
\left\{ 
\begin{array}{l}
p^{cc}\left(q_l\right) = \sum_{\begin{subarray}{c} j=0\\ j \ \mathrm{even} \end{subarray}}^{\min\left\{d_c-1,q_l\right\}}\frac{{d_c-1 \choose j}{n-d_c \choose q_l-j}}{{n-1 \choose q_l}} \\
p^{ci}\left(q_l\right) = \sum_{\begin{subarray}{c} j=0\\ j \ \mathrm{odd} \end{subarray}}^{\min\left\{d_c-1,q_l\right\}}\frac{{d_c-1 \choose j}{n-d_c \choose q_l-j}}{{n-1 \choose q_l}} \\
p^{ic}\left(q_l\right) = \sum_{\begin{subarray}{c} j=0\\ j \ \mathrm{even} \end{subarray}}^{\min\left\{d_c-1,q_l\right\}}\frac{{d_c-1 \choose j}{n-d_c \choose q_l-1-j}}{{n-1 \choose q_l-1}} \\
p^{ii}\left(q_l\right) = \sum_{\begin{subarray}{c} j=0\\ j \ \mathrm{odd} \end{subarray}}^{\min\left\{d_c-1,q_l\right\}}\frac{{d_c-1 \choose j}{n-d_c \choose q_l-1-j}}{{n-1 \choose q_l-1}} \\
\end{array} 
\right.,
\label{eq:CheckProbabilities}
\end{equation}
where $q_l$ is the average number of residual errors after the $l$th iteration.
In the considered system, it is $q_0 \leq t = t'm$, but we fix $q_0 = t = t'm$ 
to have a worst-case evaluation.

Based on these expressions, and considering the ideal assumption of a cycle-free Tanner graph, we can obtain an approximation
of the number of errors in the decoded word after the $l$th iteration.
In using this method, we do not take into account the distribution of the errors with regard to the circulant blocks weight,
that is, we consider the errors equally distributed in sets having the same cardinality for each block.
However, we have numerically verified that this approximation is largely acceptable in the considered context.
Based on these arguments, we can find $q_l$ as a function of $q_{l-1}$:
\begin{equation}
\label{eq:err_number}
q_l = t - \sum_{j=0}^{n_0-1}{\lambda_j \left[ t \cdot f^{b}\left(j,q_{l-1}\right) - \left(n-t\right) \cdot g^{b}\left(j,q_{l-1}\right)\right]}.
\end{equation}
Equation \eqref{eq:err_number} permits us to implement a recursive procedure which allows computing a waterfall threshold
by finding the maximum value $t = t_{\mathrm{th}}$ such that $\displaystyle \mathop{{\rm lim}}_{l\to \infty }\left(q_l\right)=0$.

Since different values of $t_{\mathrm{th}}$ can be found by different choices of the set of $b^{(j)}$,
we can search the maximum $t_{\mathrm{th}}$ for each combination of $b^{(j)} \in \left\{\left\lceil d_v^{(j)}/2 \right\rceil, \ldots, d_v^{(j)}-1\right\}$, with $j = 0, 1, \ldots, n_0-1$.
We will always refer to the optimal choice of the $b^{(j)}$ values in the following.

\begin{figure}[tb]
\begin{centering}
\includegraphics[keepaspectratio, width=80mm]{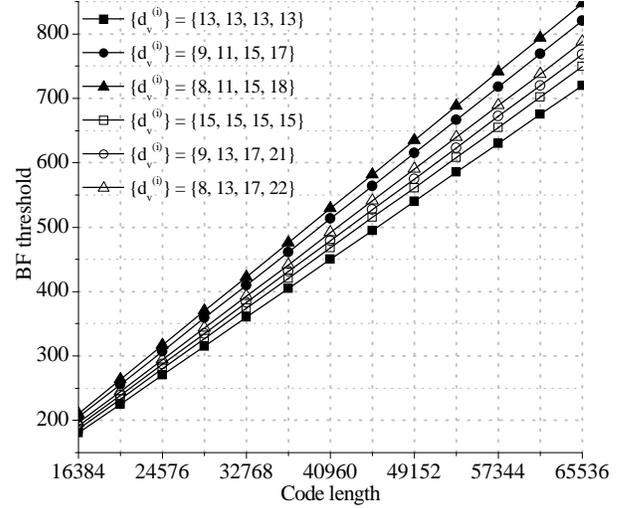}
\caption{\ac{BF} decoding threshold as a function of the code length for $n_0=4$ and several parity-check matrix column weights ($d_v$).}
\label{fig:BFthresholds}
\par\end{centering}
\end{figure}

We have used this method to compute the decoding threshold for \ac{LDPC} codes with several lengths,
$n_0 = 4$, $d_v = 13$ or $15$ and two irregular node degree profiles for each value of $d_v$ (remind that, for irregular codes, $d_v$ represents the average column weight).
The results obtained are reported in Fig. \ref{fig:BFthresholds}, where they are also compared with
the threshold values for regular codes with constant $d_v = 13$ or $15$.
These results have been obtained by considering a fixed and optimized set of decision thresholds
for the \ac{BF} decoder (that is, they do not change during iterations).
As we observe from the figure, irregular codes allow to improve the error correction
capability, coherent with the conclusion already drawn in Section \ref{sec:IrrCodes} with \ac{SPA} decoding.
It must be said that, if we consider a number of errors equal to the \ac{BF} decoding threshold,
which is computed under the hypothesis of absence of local cycles, a finite-length code with local 
cycles in its Tanner graphs does not always achieve a very low error rate under \ac{BF} decoding.
However, several improved versions of the \ac{BF} algorithm can be used, which achieve
very low residual error rates when the number of errors equals, or even overcomes, the \ac{BF} threshold \cite{Baldi2012}.
Hence, we can consider the \ac{BF} decoding threshold as a reliable estimate of the correction
capability of the codes we consider in this context.

\section{Security level}
\label{sec:SecLevel}

%The security level of \ac{QC-LDPC} code-based variants of the McEliece cryptosystem can be assessed
%by considering 
Two attack procedures mostly endanger the \ac{LDPC}-code based McEliece
cryptosystem, namely: \ac{DCA} and \ac{ISDA} \cite{Baldi2012}.
So, their \ac{WF} gives the system security level.

The target of \ac{DCA} is to recover an equivalent private key from the public key.
This can be achieved by searching for the rows of the parity-check matrix of the public code, $\mathbf{H'}$, and then
exploiting the possible sparsity of $\mathbf{H'}$ to recover $\mathbf{H}$ or to directly perform
\ac{LDPC} decoding and correct the intentional errors.
Searching for the rows of $\mathbf{H'}$ is equivalent to searching for low weight codewords
in the dual of the public code.
The matrix $\mathbf{H'}$ has average column weight $d_v' = m \cdot d_v$ and row weight $d_c' = n_0 \cdot d_v'$.
Hence, $d_v'$ is chosen high enough to make such search practically unfeasible.

The purpose of \ac{ISDA} is instead to find the error vector $\mathbf{e}$ affecting the ciphertext.
This can be accomplished through algorithms for finding low-weight codewords,
which is equivalent to decode a random linear block code.
The \ac{QC} nature of the codes facilitates this task, since each block-wise cyclically shifted 
version of a ciphertext is still a valid ciphertext.
Hence, the attacker can consider block-wise shifted versions of an intercepted ciphertext,
and search for one among as many shifted versions of the error vector.

Hence, both \ac{DCA} and \ac{ISDA} can be mounted by exploiting efficient algorithms to search
for low weight codewords in random linear block codes, and their \ac{WF} can be estimated by computing 
the minimum complexity of these algorithms.
For this purpose, we consider the approach proposed in \cite{Peters2010}. Actually, some advances 
have recently appeared in the literature \cite{May2011, Becker2012} that, however, are more 
focused on asymptotic evaluations rather than on reducing the complexity on finite length codes.
Another recent proposal in this context is ``ball collision decoding'' \cite{Bernstein2011}.
It achieves important \ac{WF} reductions asymptotically, but the improvement is negligible 
for the code lengths and security levels here of interest. 

\begin{figure}[tbh]
\begin{centering}
\includegraphics[keepaspectratio, width=80mm]{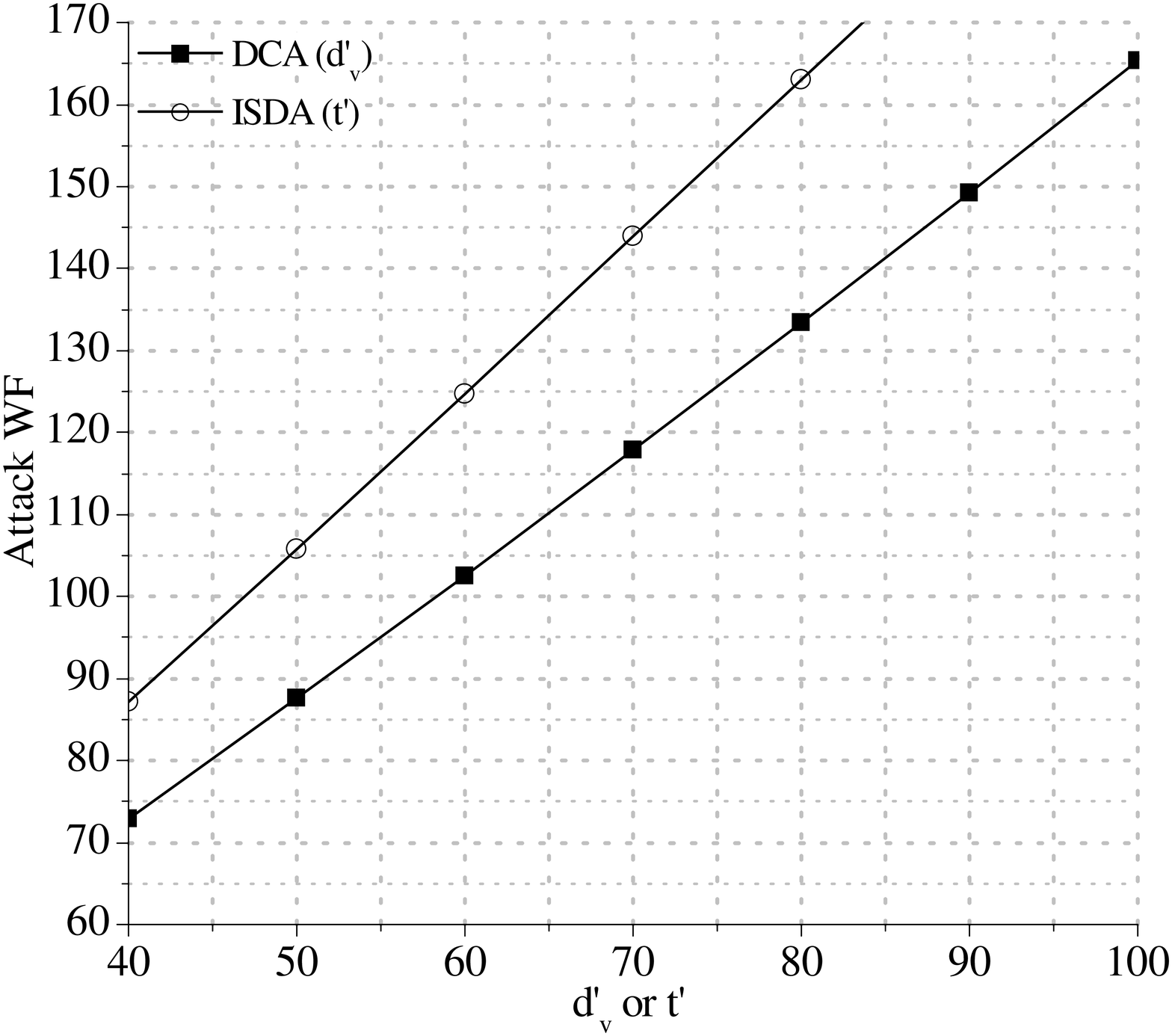}
\caption{\ac{DCA} and \ac{ISDA} \ac{WF} ($\log_2$) respectively plotted as a function of the public parity-check matrix column weight ($d_v'$)
and the number of intentional errors ($t'$), for $n_0=4$ and $n=16384$.}
\label{fig:WF}
\par\end{centering}
\end{figure}

We have computed the \ac{WF} of \ac{DCA} and \ac{ISDA}, and the
results obtained are summarized in Fig. \ref{fig:WF}.
In the figure, the abscissa reports $d_v'$ for the \ac{DCA} \ac{WF}, and $t'$ for
the \ac{ISDA} \ac{WF}.
For both attacks, the dependence of the \ac{WF} on the code length is weak, and no considerable
difference is achieved by only increasing the code length.
Hence, we have plotted the \ac{WF} for the shortest code length here considered, that is,
$n = 16384$. Using larger codes yields some increase in the \ac{WF}, which makes the value
obtained from the figure a pessimistic estimate, but without any significant deviation.

Based on the previous analysis, for a given security level and $\mathbf{H}$ column weight distribution,
the following simple procedure allows designing the system parameters:
\begin{enumerate}
\item the values of $d_v'$ and $t'$ needed for achieving the desired security level are obtained from Fig. \ref{fig:WF};
\item the value of $m$ is computed as $d_v' / d_v$;
\item the number of intentional errors to correct is computed as $t = \left\lceil m \cdot t' \right\rceil$;
\item the code length is found from Fig. \ref{fig:BFthresholds}, such that the corresponding \ac{BF} threshold overcomes $t$.
\end{enumerate}

We notice that using an irregular matrix $\mathbf{Q}$ avoids the need to increase $m$ up to $\left\lceil m \right\rceil$, thus keeping the error propagation effect of
$\mathbf{Q}$ as small as possible.
This increases the efficiency of the system, since $t$ and, hence, the code length are kept to their minimum.
We remind, however, that $m$ must be a multiple of $1 / {n_0^2}$, hence we must approximate it
to the smallest multiple of $1 / {n_0^2}$ greater than or equal to $d_v' / d_v$.

\section{Design examples}
\label{sec:Examples}

Let us suppose to need $100$-bit security.
From Fig. \ref{fig:WF} we obtain $d_v' = 59$ and $t' = 47$.
If we focus on $d_v = 13$ for the private code, it results $m = 4.5625$
(approximated to a multiple of $1/n_0^2 = 1/16$).
Then, $t = \left\lceil m \cdot t' \right\rceil = 215$.
From Fig. \ref{fig:BFthresholds} we find that a regular code with $d_v=13$
and $n = 20480, p = 5120$ has a \ac{BF} threshold equal to $225$; hence, it is able
to correct all intentional errors.
This yields a key size of $15360$ bits.
By looking at the irregular code with $d_v = 13$ and degree profile $\left\{ 8, 11, 15, 18 \right\}$,
we obtain that the same \ac{BF} decoding threshold is achieved for $n = 17524, p = 4381$, and
the key size becomes $13143$ bits, which is a $15 \%$ reduction with respect to the regular code.

If we want to achieve $160$-bit security, we obtain from Fig. \ref{fig:WF}
that $d_v' = 97$ and $t' = 79$ are needed.
By still considering $d_v = 13$, we obtain $m = 7.4375$ (with the same approximation as before).
It follows that $t = 588$.
From Fig. \ref{fig:BFthresholds}, we see that a regular code with $d_v=13$
and $n = 54616, p = 13654$ has a \ac{BF} threshold equal to $600$, which is enough
to correct all intentional errors. The corresponding key size is $40962$ bits.
If we use an irregular code with degree profile $\left\{ 8, 11, 15, 18 \right\}$,
the same \ac{BF} decoding threshold is achieved for $n = 46448, p = 11622$, that is, a key
size of $34866$ bits.
Hence, using an irregular code results in a public key size reduction of about $15 \%$ 
also in this case.

\section{Conclusion}
\label{sec:Conclusion}

We have studied the use of irregular codes in the McEliece cryptosystem based on \ac{LDPC} codes.
We have considered \ac{QC-LDPC} codes with irregular degree profiles, and verified, through numerical
simulations and theoretical tools, that they achieve better error correction performance than regular 
codes also in this context.
This reflects into a more efficient cryptosystem, with a public key size reduction in the order of
$15 \%$ with respect to the version using regular codes.

Future work will concern the evaluation of the implementation cost for the proposed solutions \cite{Sklavos2010}.
% and their usage in the framework of practical secure systems \cite{Sklavos2007}.

\newcommand{\BIBdecl}{\setlength{\itemsep}{0.01\baselineskip}}



\begin{thebibliography}{10}
\providecommand{\url}[1]{#1}
\csname url@samestyle\endcsname
\providecommand{\newblock}{\relax}
\providecommand{\bibinfo}[2]{#2}
\providecommand{\BIBentrySTDinterwordspacing}{\spaceskip=0pt\relax}
\providecommand{\BIBentryALTinterwordstretchfactor}{4}
\providecommand{\BIBentryALTinterwordspacing}{\spaceskip=\fontdimen2\font plus
\BIBentryALTinterwordstretchfactor\fontdimen3\font minus
  \fontdimen4\font\relax}
\providecommand{\BIBforeignlanguage}[2]{{%
\expandafter\ifx\csname l@#1\endcsname\relax
\typeout{** WARNING: IEEEtran.bst: No hyphenation pattern has been}%
\typeout{** loaded for the language `#1'. Using the pattern for}%
\typeout{** the default language instead.}%
\else
\language=\csname l@#1\endcsname
\fi
#2}}
\providecommand{\BIBdecl}{\relax}
\BIBdecl

\bibitem{McEliece1978}
R.~J. McEliece, ``A public-key cryptosystem based on algebraic coding theory.''
  \emph{DSN Progress Report}, pp. 114--116, 1978.

\bibitem{Bernstein2008}
D.~J. Bernstein, T.~Lange, and C.~Peters, ``Attacking and defending the
  {McEliece} cryptosystem,'' in \emph{Post-Quantum Cryptography}, ser. Lecture
  Notes in Computer Science.\hskip 1em plus 0.5em minus 0.4em\relax Springer
  Verlag, 2008, vol. 5299, pp. 31--46.

\bibitem{Baldi2008}
M.~Baldi, M.~Bodrato, and F.~Chiaraluce, ``A new analysis of the {McEliece}
  cryptosystem based on {QC-LDPC} codes,'' in \emph{Security and Cryptography
  for Networks}, ser. Lecture Notes in Computer Science.\hskip 1em plus 0.5em
  minus 0.4em\relax Springer Verlag, 2008, vol. 5229, pp. 246--262.

\bibitem{Baldi2012}
\BIBentryALTinterwordspacing
M.~Baldi, M.~Bianchi, and F.~Chiaraluce, ``Security and complexity of the
  {McEliece} cryptosystem based on {QC-LDPC} codes,'' \emph{IET Information
  Security}, 2012, in press. [Online]. Available:
  \url{http://arxiv.org/abs/1109.5827}
\BIBentrySTDinterwordspacing

\bibitem{Baldi2013}
------, ``Optimization of the parity-check matrix density in {QC-LDPC}
  code-based {McEliece} cryptosystems,'' in \emph{Proc. {IEEE ICC} 2013 -
  Workshop on Information Security over Noisy and Lossy Communication Systems},
  Budapest, Hungary, Jun. 2013.

\bibitem{Misoczki2012}
\BIBentryALTinterwordspacing
R.~Misoczki, J.-P. Tillich, N.~Sendrier, and P.~S. L.~M. Barreto. (2012)
  {MDPC-McEliece}: New {McEliece} variants from moderate density parity-check
  codes. [Online]. Available: \url{http://eprint.iacr.org/2012/409}
\BIBentrySTDinterwordspacing

\bibitem{Biasi2012}
\BIBentryALTinterwordspacing
F.~P. Biasi, P.~S. L.~M. Barreto, R.~Misoczki, and W.~V. Ruggiero. (2012)
  Scaling efficient code-based cryptosystems for embedded platforms. [Online].
  Available: \url{http://arxiv.org/abs/1212.4317}
\BIBentrySTDinterwordspacing

\bibitem{Richardson2001}
T.~J. Richardson and R.~L. Urbanke, ``The capacity of low-density parity-check
  codes under message-passing decoding,'' \emph{{IEEE} Trans. Inform. Theory},
  vol.~47, no.~2, pp. 599--618, Feb. 2001.

\bibitem{Paolini2009}
E.~Paolini and M.~Chiani, ``Construction of near-optimum burst erasure
  correcting low-density parity-check codes,'' \emph{{IEEE} Trans. Commun.},
  vol.~57, no.~5, pp. 1320--1328, May 2009.

\bibitem{Baldi2012b}
M.~Baldi, M.~Bianchi, G.~Cancellieri, and F.~Chiaraluce, ``Progressive
  differences convolutional low-density parity-check codes,'' \emph{{IEEE}
  Commun. Lett.}, vol.~16, no.~11, pp. 1848--1851, Nov. 2012.

\bibitem{Baldi2012c}
M.~Baldi, G.~Cancellieri, and F.~Chiaraluce, ``Interleaved product {LDPC}
  codes,'' \emph{{IEEE} Trans. Commun.}, vol.~60, no.~4, pp. 895--901, Apr.
  2012.

\bibitem{Baldi2012a}
M.~Baldi, M.~Bianchi, and F.~Chiaraluce, ``Coding with scrambling,
  concatenation, and {HARQ} for the {AWGN} wire-tap channel: A security gap
  analysis,'' \emph{IEEE Trans. Inf. Forensics Security}, vol.~7, no.~3, pp.
  883--894, Jun. 2012.

\bibitem{Monico2000}
C.~Monico, J.~Rosenthal, and A.~Shokrollahi, ``Using low density parity check
  codes in the {M}c{E}liece cryptosystem,'' in \emph{Proc. {IEEE} International
  Symposium on Information Theory (ISIT 2000)}, Sorrento, Italy, Jun. 2000, p.
  215.

\bibitem{Baldi2007ISIT}
M.~Baldi and F.~Chiaraluce, ``Cryptanalysis of a new instance of {McEliece}
  cryptosystem based on {QC-LDPC} codes,'' in \emph{Proc. {IEEE} International
  Symposium on Information Theory (ISIT 2007)}, Nice, France, Jun. 2007, pp.
  2591--2595.

\bibitem{Baldi2007ICC}
M.~Baldi, F.~Chiaraluce, R.~Garello, and F.~Mininni, ``Quasi-cyclic low-density
  parity-check codes in the {McEliece} cryptosystem,'' in \emph{Proc. IEEE
  International Conference on Communications ({ICC} 2007)}, Glasgow, Scotland,
  Jun. 2007, pp. 951--956.

\bibitem{Luby2001a}
M.~Luby, M.~Mitzenmacher, M.~Shokrollahi, and D.~Spielman, ``Improved
  low-density parity-check codes using irregular graphs,'' \emph{{IEEE} Trans.
  Inform. Theory}, vol.~47, no.~2, pp. 585–--598, Feb. 2001.

\bibitem{Baldi2011a}
M.~Baldi, F.~Bambozzi, and F.~Chiaraluce, ``On a family of circulant matrices
  for quasi-cyclic low-density generator matrix codes,'' \emph{{IEEE} Trans.
  Inform. Theory}, vol.~57, no.~9, pp. 6052--6067, Sep. 2011.

\bibitem{Johnson2010}
S.~J. Johnson, \emph{Iterative Error Correction}.\hskip 1em plus 0.5em minus
  0.4em\relax New York, NY: Cambridge University Press, 2010.

\bibitem{Hagenauer1996}
J.~Hagenauer, E.~Offer, and L.~Papke, ``Iterative decoding of binary block and
  convolutional codes,'' \emph{{IEEE} Trans. Inform. Theory}, vol.~42, no.~2,
  pp. 429--445, Mar. 1996.

\bibitem{Baldi2009a}
M.~Baldi, G.~Cancellieri, and F.~Chiaraluce, ``Finite-precision analysis of
  demappers and decoders for {LDPC}-coded {M-QAM}-systems,'' \emph{{IEEE}
  Trans. Broadcast.}, vol.~55, no.~2, pp. 239--250, Jun. 2009.

\bibitem{Gallager1963}
R.~G. Gallager, \emph{Low-density parity-check codes}.\hskip 1em plus 0.5em
  minus 0.4em\relax M.I.T. Press, 1963.

\bibitem{Peters2010}
C.~Peters, ``Information-set decoding for linear codes over {$F_q$},'' in
  \emph{Post-Quantum Cryptography}, ser. Lecture Notes in Computer
  Science.\hskip 1em plus 0.5em minus 0.4em\relax Springer Verlag, 2010, vol.
  6061, pp. 81--94.

\bibitem{May2011}
A.~May, A.~Meurer, and E.~Thomae, ``Decoding random linear codes in
  {$O(2^{0.054n})$},'' in \emph{ASIACRYPT 2011}, ser. Lecture Notes in Computer
  Science.\hskip 1em plus 0.5em minus 0.4em\relax Springer Verlag, 2011, vol.
  7073, pp. 107--124.

\bibitem{Becker2012}
A.~Becker, A.~Joux, A.~May, and A.~Meurer, ``Decoding random binary linear
  codes in $2^{n/20}$: How 1 + 1 = 0 improves information set decoding,'' in
  \emph{EUROCRYPT 2012}, ser. Lecture Notes in Computer Science.\hskip 1em plus
  0.5em minus 0.4em\relax Springer Verlag, 2012.

\bibitem{Bernstein2011}
D.~J. Bernstein, T.~Lange, and C.~Peters, ``Smaller decoding exponents:
  ball-collision decoding,'' in \emph{CRYPTO 2011}, ser. Lecture Notes in
  Computer Science.\hskip 1em plus 0.5em minus 0.4em\relax Springer Verlag,
  2011, vol. 6841, pp. 743--760.

\bibitem{Sklavos2010}
N.~Sklavos, ``On the hardware implementation cost of crypto-processors
  architectures,'' \emph{Information Security Journal}, vol.~19, no.~2, pp.
  53--60, Apr. 2010.

\end{thebibliography}
\end{document}